\documentclass[aps,prl,twocolumn,superscriptaddress,groupedaddress]{revtex4}  % for review and submission
\usepackage{graphicx}  % needed for figures
\usepackage{dcolumn}   % needed for some tables
\usepackage{bm}        % for math
\usepackage{amssymb}   % for math
\usepackage{ marvosym }
\usepackage[font=small,skip=4pt]{caption}
\begin{document}

%\title{High-Resolution Hard X-ray Holography by Deconfined Phase-Shifting 3D References by Atomic Layer Deposition} %ALD-Coated }
\title{High-Resolution Hard X-ray Holography by Unconfined Atomic Layer Deposited Phase-Shifting 3D References} %ALD-Coated }
%\title{High-Resolution Hard X-ray Lensless Imaging by Unconfined ALD-coated Holographic References} 

\affiliation{Universit{\"a}t Z{\"u}rich, Z{\"u}rich, Switzerland}
\affiliation{Diamond Light Source Ltd, Oxfordshire, United Kingdom}
\affiliation{Paul Scherrer Institute, Villigen, Switzerland}
\affiliation{Southhampton University, South Hampton, United Kingdom}

\author{M.T.~Saliba $^{\ast}$} \affiliation{Universit{\"a}t Z{\"u}rich, Z{\"u}rich, Switzerland}\affiliation{Diamond Light Source Ltd, Oxfordshire, United Kingdom}\thanks{mirna.saliba@diamond.ac.uk}
\author{J.~Bosgra} \affiliation{Paul Scherrer Institute, Villigen, Switzerland}
\author{C.~Rau} \affiliation{Diamond Light Source Ltd, Oxfordshire, United Kingdom}
\author{C.~David} \affiliation{Paul Scherrer Institute, Villigen, Switzerland}
\author{A.D.~Parsons} \affiliation{Diamond Light Source Ltd, Oxfordshire, United Kingdom}
\author{U.H.~Wagner} \affiliation{Diamond Light Source Ltd, Oxfordshire, United Kingdom}
\author{P.~Thibault}\affiliation{Diamond Light Source Ltd, Oxfordshire, United Kingdom} \affiliation{Southhampton University, South Hampton, United Kingdom}

%
% list_of_visitor_addresses_r2.tex            24 March 2010
%  available symbols are:
%  $\ast, \dag, \ddag, \S, \P, $\|$, $\ast\ast$, \dag\dag, \ddag\ddag ,\#
%
\vskip 0.25cm
      
%\date{\today}
\date{August 31, 2016}

\begin{abstract}
We demonstrate high-resolution non-iterative holographic coherent diffraction imaging with hard X-rays using a novel phase-shifting reference, fabricated by atomic layer deposition to produce nano-sharp 3D structure. The method surpasses the limitations associated with absorbing substrates predominantly employed in soft X-ray holography using extended, customized and point-source references. The unconfined experimental setup relaxes the technical constraints, allows effective data correction, and enables independent sample translation and rotation for data averaging and tomography. Applicable to single-shot measurements, phase and amplitude reconstructions of samples are retrieved with single-pixel resolution and differential contrast by simple non-iterative computation. \end{abstract}

\maketitle

In the interest to unveil structural information of matter at the nanometer scale and beyond, there is an endless quest to innovate new techniques that provide images and maps of a probed sample. From the first X-ray radiograph obtained by Wilhelm R{\"o}ntgen in 1895, to the latest developments in microscopy and crystallography, where carbon atoms in a graphene layer can be \textit{seen} and protein structure can be visualised on the atomic level  \cite{chapmanprotein}, the field of imaging and microscopy is forever evolving. Today, it is the caliber of structural information of a probed sample in a static or dynamic setting that makes the field of imaging very competitive and gives rise to a vast variety of techniques. 
When probing radiation-sensitive samples and ultrafast molecular processes, the aim is to capture a snapshot of a state or event before the sample changes or disintegrates \cite{spencediffdest,mimivirus} and reconstruct the data in a short time-frame to validate the results or update the measurement. This triggers the need for a single-shot imaging method, unreliant of image-forming lenses, that can directly and non-iteratively retrieves  complex-valued reconstructions with high resolution, all the while being robust, self-subsistent, and computationally inexpensive.
This is especially motivated by the development of free-electron-lasers that provide highly brilliant and coherent X-rays in a single femtosecond pulse \cite{chapxfelsoft,biopulsesneutze}.

Holography is an imaging technique that intrinsically preserves the phase information encoded by virtue of a reference wave \cite{gabor}. It is a product of the superposition of a scattered object wave with the reference wave and involves non-iterative reconstruction of the complex wave-field of the sample by solving the Kirchoff-Fresnel integral. Fourier transform holography \cite{mcnulty,eisebit1} (FTH) and Holography with extended reference by autocorrelation linear differential operation \cite{guizarmain,heraldozhucheese} (HERALDO) emerged as simple and effective methods to disentangle the twin images (inherent to in-line holography) by means of an off-axis reference such as a pinhole \cite{eisebit1,multsoftfth}, an extended aperture with sharp corners \cite{manuthesis}, or a uniformly redundant array (URA) \cite{URA}. Another development involved the use of a customizable reference and applying iterative linear reconstruction using Fourier transforms (ILRUFT) methods \cite{ilruft}. Mask-assisted phase-amplitude retrieval has been demonstrated numerically and experimentally using visible light, where the sample is placed within a rectangular aperture or in the vicinity of an opaque rectangular mask \cite{podorov, podorovmask}. The method termed Differential Fourier Holography (DFH) was also demonstrated numerically for the investigation of simulated three dimensional objects \cite{podorov3d}. With FTH and HERALDO, the reconstruction is obtained by a single Fourier transform owing to the the Wiener-Khinchin theorem and the sifting property of the delta function where the transmission function of a pinhole or an extended reference is approximated or differentially reduced to a single or multiple delta functions \cite{guizarmain}. The attractive features of HERALDO are: robustness against noise, drift, and data imperfections; simple preparation and uncomplicated experimental implementation; and direct reconstruction of single-shot data by a single computational step.

We have developed a holographic diffraction imaging technique particularly tailored for hard X-rays but generally transferable to all coherent radiation. The main feature is the use of a novel reference 3D structure with phase-shifting properties that surpasses the limitations associated with the holographic performance and fabrication of absorbing masks. With nano-layer edges produced by innovative nano-fabrication based on atomic layer deposition, the upgraded reference design elevates the resolution and contrast of the imaging system beyond the fabrication capacity of electron/ion-beam milling. By simple non-iterative computation of a single-shot dataset, high-resolution reconstructions of a specimen are obtained in amplitude and phase. Due to the unique reference structure, reconstructions portray differential phase contrast and edge-contrast amplification. In addition, we implemented a flexible experimental design that involves the unconfinement of the sample plane and the reference plane mitigating many of the experimental constraints and allowing translation and rotation of the two planes independently. This leads to enhanced data refinement and background filtering, better statistics by sample shifting, accommodating a large set of samples, and facilitating tomographic  measurements \cite{tomo}.
\\Until now, holographic methods such as FTH \cite{eisebit1,femtoseq,multsoftfth}, HERALDO \cite{heraldofemto,heraldointegralcrosswire,heraldowindocapo}, DFH \cite{podorov,podorovmask,podorov3d} and ILRUFT \cite{ilruft} have been mostly implemented using visible light, soft X-ray radiation and less applications with hard X-rays \cite{hard1,hard2}. The required thickness of an absorbing holographic mask suitable for soft X-rays is generally in the sub-micron range. On the other hand, with strongly penetrating hard X-rays, the required thickness for full absorption is in the sub-millimeter range.
It is a straightforward task to mill an aperture using a focused ion beam (FIB) in a sub-micron membrane of silicon nitride, for instance, with sub-100nm edge resolution, yet this is virtually impossible for a 0.5 mm substrate. 
The ease of fabrication facilitates the application of HERALDO with soft x-rays and is indicative to why the predominantly used type of holographic reference is that of a milled aperture in an absorbing substrate \cite{eisebit1,femtoseq,heraldozhucheese,heraldointegralcrosswire} with some exceptions \cite{podorovmask,enders}. 
However, hard X-ray radiation has the virtue of having high penetration depth and low diffuse scattering and phase sensitivity which means it provides depth information about thick samples, thus serving as a powerful tool in medical imaging as well as archeology, semiconductors and the material sciences. 

Therefore, holographic imaging using extended references in the hard X-ray regime requires a novel reference design that alleviates the fabrication constraints yet delivers high resolution and contrast. This is accomplished by adopting a phase-shifting holographic mask \cite{enders,enders2} instead of an absorbing/transmissive mask. In either case, the reference wave is a result of the boundary waves generated by the sharp transition in complex transmissivity (amplitude and phase) across the edge of the aperture. The height of the transition (difference in absorption and phase shift) dictates the image contrast, and the width of the transition (edge sharpness) governs the resolution. A suitable phase-shifting mask for hard X-rays at 8 keV - 20 keV is a silicon pillar in the order of 10 $\mu m$ thickness with a phase of $\pi$ on a transmissive substrate with a phase of $0\pm 2\pi$. With negligible absorption across the mask, it is the sharp transition in phase that generates boundary waves across the edge of the pillar representing the reference wave. The reference transmission function is proportional to $|\tau_r - \tau_s |$, where $\tau_r$ and $\tau_s$ are the complex transmissivities of the reference structure and the substrate respectively. Compared to an absorbing mask, the implication of this is a fourfold amplification of the reference signal intensity \cite{manuthesis,enders}, as $|\tau_r - \tau_s |^2 = |e^{i\pi} - e^{i0} |^2 =4$, neglecting absorption. The signal strength of the object reconstruction, resulting from the cross-correlation between the reference function and the object function, is also amplified by a factor of two. This proves despite the negligible absorption of the reference mask, the $\pi$ phase shift is independently responsible for high contrast in holographic imaging, which bodes well for applications with low-absorption phase samples and low-density weakly scattering samples such as biological samples.

The fabrication of phase-shifting 3D reference structures relies on nanolithography by electron beam exposure and crystal etching along the direction of the crystal lattice planes. Such techniques are highly reproducible and routinely achieve sub-100nm edge sharpness. In theory, atomically-sharp edges can be obtained using a perfect crystal and precision alignment of the mask etching with respect to the crystallographic orientation.
Yet, within the practical capacity, a standardly etched silicon pillar will exhibit some edge roughness at the nanometer scale in addition to occasional etching imperfections and lingering debris. The application of references fabricated by such a technique was demonstrated in  \cite{mirna} with a resolution reaching 100 nm and some artefacts due to etching imperfections.  
Since the image resolution is ultimately limited by the sharpness of the reference edges, the refinement of reference structures beyond the etching limit is highly valuable. Here, we have employed an innovative fabrication technique to produce phase-shifting reference structures with high-resolution edges that exceeds the quality of basic silicon etching nano-fabrication. 
\begin{figure}[!b]
	\includegraphics[scale=0.465]{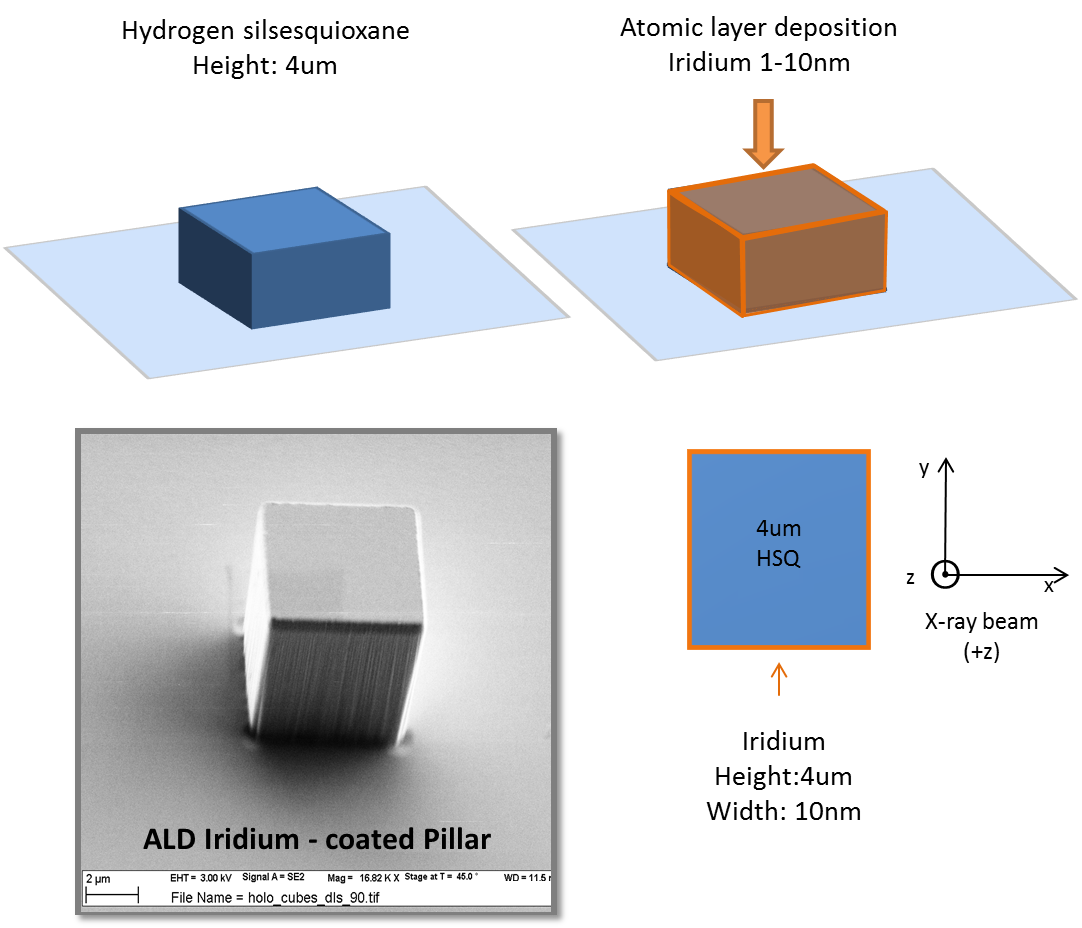}
	\caption{{\small Schematic of ALD fabrication. (lower left) SEM image of ALD iridium-coated HSQ pillar. Scale bar: 2$\mu m$}}
	\label{ald_fab}
\end{figure} 
\setlength{\belowcaptionskip}{20pt}
\begin{figure*}[!t]
	\vspace{-0.5cm}
	\includegraphics[scale=0.7]{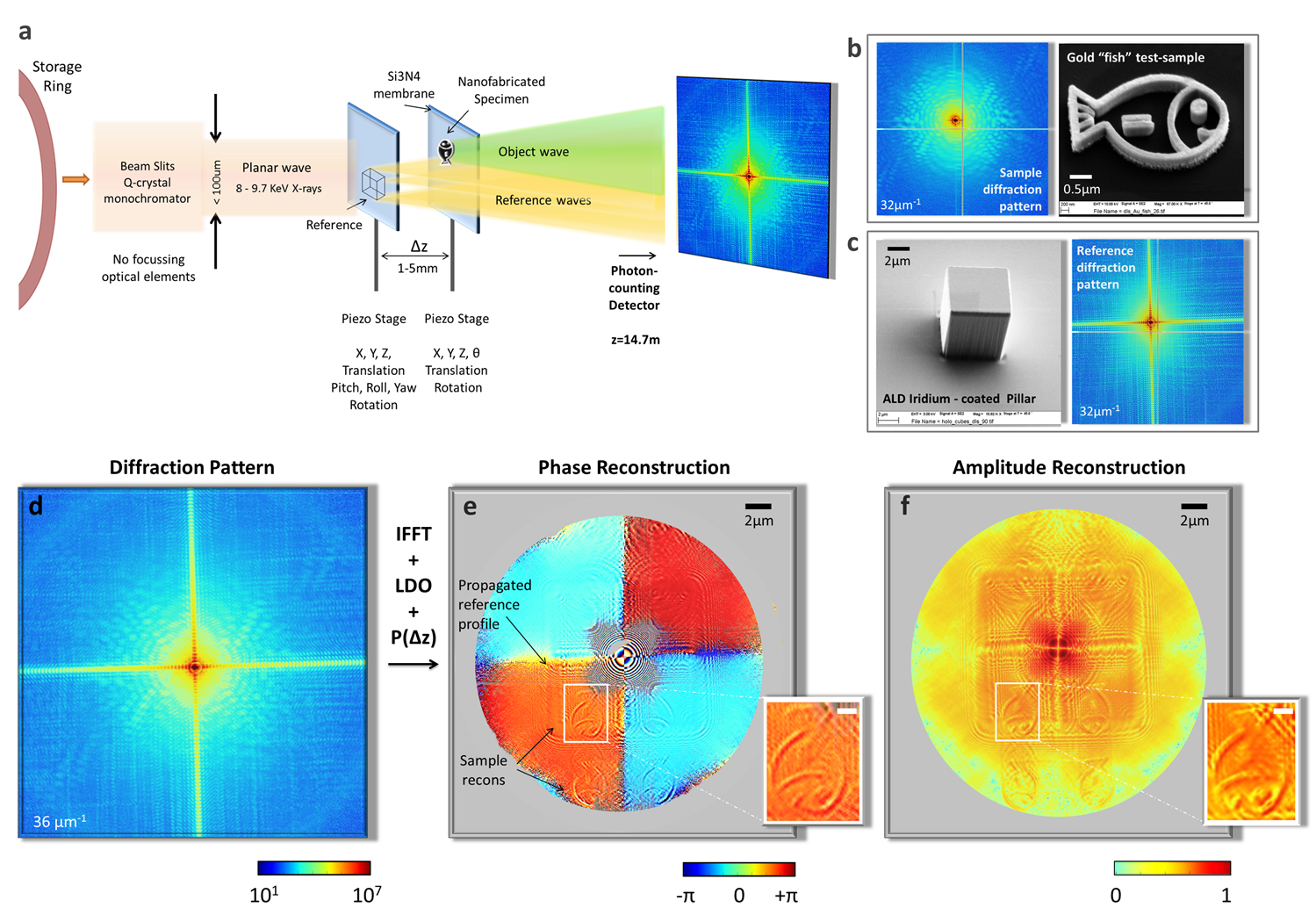}
	\caption{a. Schematic of the experimental setup; b,c. individual diffraction pattern of sample and reference, respectively, and SEM images; d. holographic diffraction pattern; e,f. reconstructed phase and amplitude respectively by IFFT, LDO, and near-field propagation P($\Delta z$). (e,f)(lower right) zoomed-in sample reconstruction. Scale bar: 500nm.}
	\label{schem}
\end{figure*}

\vspace{-10pt}
The fabrication upgrade involves coating a hydrogen silsequioxane (HSQ) pillar with a 10 nm layer of iridium by atomic layer deposition (ALD). ALD allows the deposition of one iridium atom upon the other in a very structured manner such that the process results in a pillar enveloped with a sharp iridium nano-edge (Figure \ref{ald_fab}). The sharpness of the iridium edge will contribute to high resolution reconstructions while the electron-dense material contributes increased contrast compared to a bare silicon pillar. Furthermore, when viewing the ALD-coated pillar from the perspective of the incoming X-ray beam, the reference structure exhibits the shape of a square wireframe as opposed to the bare silicon pillar which appears as a full-area square. With the face of the pillar aligned orthogonally to the propagation direction of the beam, the X-rays \textit{see} a square area of light material (HSQ) with minimal absorption and modest phase shift, surrounded by a 10 nm frame of electron-dense material (iridium) that is laterally thin but longitudinally thick, thus contributing both considerable absorption and $\pi$ phase shift along the propagation direction. 
\\
\indent The specific design of the phase-shifting mask holds several advantages for the holographic system. Since, contrary to the absorbing mask, the reference structure lies on a transmissive substrate, this allows to physically decouple the sample and the reference. In mainstream FTH and HERALDO experiments, both the reference and the sample are fabricated monolithically \cite{monolithic} (on the same substrate) as the surrounding substrate is absorbing \cite{femtoseq,ilruft}. In comparison, in the case of the phase-shifting reference on a transmissive substrate, the incident parallel beam can illuminate the reference structure and propagate through the substrate unadulteratedly to impinge on a laterally offset sample placed in a separate plane behind the reference. This does not disturb the holographic mechanism as the sample images are obtained by near-field propagation of the cross-correlation reconstructions by a few millimeters, which is an unsubstantial distance for a far-field diffraction setup at such X-ray wavelengths. The two components, being unconfined from one another, enjoy new degrees of freedom through translation and rotation. This design relaxes many experimental preconditions, opens new possibilities for tomographic imaging \cite{tomo}, and allows scanning through a large set of samples without unmounting or readjusting. It also leads to optimised measurements by laterally shifting the components for optimal exposure and coherence and satisfying the oversampling and holographic separation conditions \cite{manuthesis}. In addition, it enables data refinement by subtraction of the beam, sample, or reference autocorrelations individually to correct for background artefacts and enhance the signal of the sample reconstruction. 
\\
\indent The experiment was performed at the I13-1 \textit{Coherence} Beamline at Diamond Light Source, UK. The experimental hutch is located at 250 meters from the synchrotron source to provide high coherence in the order of 300 $\mu m$ vertically and 150 $\mu m$ horizontally. As illustrated in Figure \ref{schem}(a), far-field holographic diffraction measurements were taken at 9.1 keV photon energy, using a flat and uniform monochromatic beam illumination of 100$\mu m$ diameter that was shaped using beam-forming slits passing through a double-crystal monochromator, complex refractive lenses, and KB mirror. The sample and reference planes were mounted on individual on nano-precision piezo-electric stages separated by 3.5 mm in the propagation direction. 
By making use of the unconfined setup, additional measurements were obtained where the sample and reference was iteratively shifted by 2$\mu m$ steps in (x,y) to find the optimal position in terms of flux, coherence, and minimal overlap in the reconstruction. The data was recorded using a Merlin photon counting detector with 515$\times$515 pixels  and 55$\mu m$ pixel size placed at 14.7m from the sample. The total number of photons at the detector reached 2.45$\times 10^9$ counts per second with exposure time between 100 sec and 500 sec. 

To further refine the data, for every setup three additional measurements of the diffraction patterns were obtained of the beam, the sample, and the reference individually, at fixed energy and beam parameters. This was accomplished by translating each of the reference and the sample structures by permutation in and out of the beam path while transmitting through the surrounding substrate. These datasets are then used for background filtering in either Fourier space or real space. The measurements are also informative of the shape and structure of the reference and sample as well as beam artefacts. In addition, the signal strength and maximum scattering angle of the individual diffraction patterns indicate the expected image contrast and resolution respectively. By Fourier transform, we compute the individual autocorrelations and use that information to adjust the experimental parameters as needed. In a sense, the process providing live feedback represents an optimal experimental procedure that involves "smart" and informed measurement and economic use of beam-time, rather than "blind" measurement relying on post-processing through time-consuming algorithms. Such an advantage is unattainable with a monolithic reference-sample mask. 

The reconstruction procedure is composed of three computational tools. The first is the inverse Fourier transform (IFFT) to obtain the autocorrelation $\psi \otimes \psi$ of the total transmissivity of the reference $r(x,y)$ and the object $o(x,y)$, where $\psi(x,y)=r(x,y)+o(x,y)$. The second fundamental element is differentiation that reduces the square reference function to four delta peaks which cross-correlate with the object function to produce the object reconstructions. Differentiation can be performed in real space or reciprocal space. The real-space linear differential operator (LDO) \cite{guizarmain} involves a double directional derivative applied to IFFT$\{I(p,q)\}$, where $I(p,q)$ is the measured intensity distribution, leading to the differentiated autocorrelation $\mathcal{L}^2 \{\psi \otimes \psi\}$. Equivalent to differentiation in real space is a linear multiplicative function in q-space such that the reconstruction is obtained by $\mathcal{F}^{-1} \{(2\pi)^2 q.p. I(q,p)\}$ where the filter is applied directly to the intensity distribution and the IFFT is processed at the end. In both cases, the direction of differentiation is tailored according to the directions of the square sides and the angle between them. This can be readily deduced by inspecting the angle of the rectangular fringes in the diffraction pattern or more clearly by the reference autocorrelation. The third element of the reconstruction procedure, inherent to the unconfined setup, is near-field propagation to bring the cross-correlation reconstructions into focus at a distance $+\Delta z$, and similarly their twin images at $-\Delta z$. 
\begin{figure*}[!t]
	\vspace{-0.4cm}
	\includegraphics[scale=0.375]{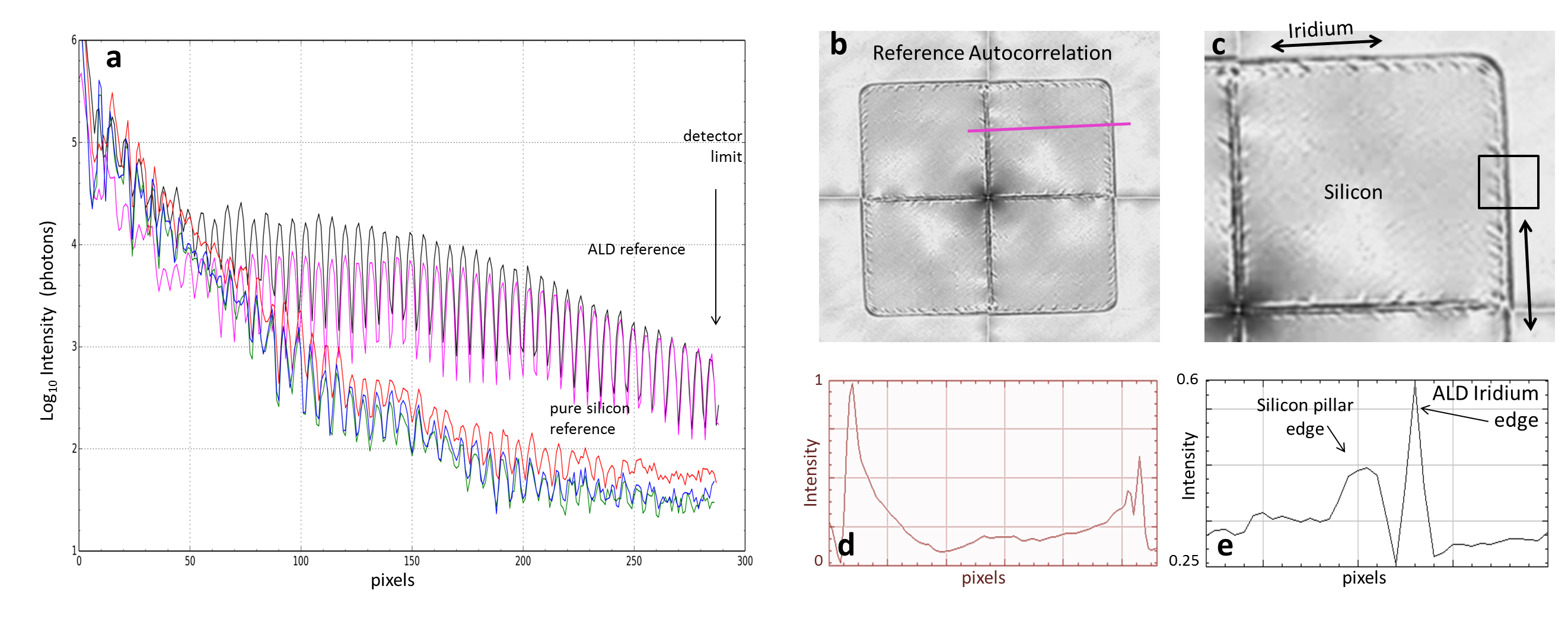}
	\vspace{-0.3cm}
	\caption{{\small a. Plot of intensity distribution comparing diffraction fringes for ALD reference at 500sec (black), 200sec (pink) exposure, and for silicon pillar at 200sec (red), 100sec (blue), and 50sec (green) exposure; b. autocorrelation reconstruction of ALD iridium-coated HSQ pillar; c. zoom-in section of b; d. profile plot of pink line in b; e. profile plot of inset in c.}}
	\label{refmod}
\end{figure*}

Figure \ref{schem}(b) and (c) show the individual diffraction patterns of the sample and reference respectively with the corresponding SEM images. The holographic diffraction pattern shown in Figure \ref{schem}(d) is obtained by illuminating a 6$\times$6$\times$3.9$\mu m^3$ ALD iridium/HSQ wireframe reference and a cartoon-fish sample with 340 nm gold thickness and 4$\mu m$ length, at 200 sec exposure time. The horizontal and vertical rectangular sinc modulations are distributed across the entire detector with high intensity and visibility, superposed with the radially distributed intense speckle pattern of the gold sample.  Dissimilar to the diffraction pattern of a two-dimensional slit aperture, we observe fine rectangular modulations spread across the detector attributed to the 3D cuboid structure.
The phase and amplitude reconstructions are shown in Figure (e) and (f) respectively. The central region of the reconstructed area contains the reference autocorrelation appearing as four adjacent copies of the square structure of the reference. The sharp linear Fresnel fringes around the reference autocorrelation are due to near-field propagation of the dense iridium edges. Four images of the sample appear in-focus on the bottom side, as the top twin images are out-of-focus. The sample reconstructions are a product of the cross-correlation between the sample function and the delta functions at the four corner positions of the square reference, as a result of the directional differential filter. Each sample image occurs at the virtual position of the reference corner. In Figure \ref{schem}(e), the phase is alternating across the four quadrants of the reconstructed field of view, corresponding to the phase-shift value of the reference and its conjugate.  
The magnified sample reconstructions shown in Figure \ref{schem}(e,f) exhibit high contrast and resolution both in amplitude and phase. The 150nm outline of the gold fish is well-resolvable. The detail structure of the tail ranging from 50nm to 100nm and the 40nm gap between the two fins and are especially distinguishable in the phase.

The sample is positioned such that separation conditions are not entirely satisfied. We find that even when the sample reconstructions overlap with the reference autocorrelation, the intensity of the sample image is not overpowered by the reference autocorrelation but is sufficiently high to be distinguished in the overlap area. This is due to the fact that the inner reference area is that of a 3.9$\mu m$-thick silicon which contributes almost full transmission and low phase-shift. This proves that the image contrast is mainly a product of the $\pi$-phase-shifting iridium edge, augmented by virtue of the contrast amplification inherent to the phase-shifting holographic mechanism. Despite the fine width of the nano-wireframe, the fact that it is laterally extended to a perimeter of a $6\mu m$ square and longitudinally extended by $3.9\mu m$ translates into high reference signal strength to produce sample reconstructions with sufficiently high contrast and signal-to-noise ratio (SNR). The advantage of the square reference geometry is that the mean squared error of the reconstructions (inverse of the SNR) is independent of the area or sharpness of the reference \cite{manuthesis}. This has an important implication in relation to the point reference or pinhole scenario that suffers from the resolution-contrast trade-off \cite{eisebit1,URA}, where the sharper the point the lower the contrast and SNR but the higher the resolution. Contrarily, with the square reference, even an infinitely sharp reference edge does not affect the contrast and SNR at all - it only enhances resolution - as it does not diminish the boundary wave generated at the edge transition.
\begin{figure*}[!t]
	\includegraphics[scale=0.447]{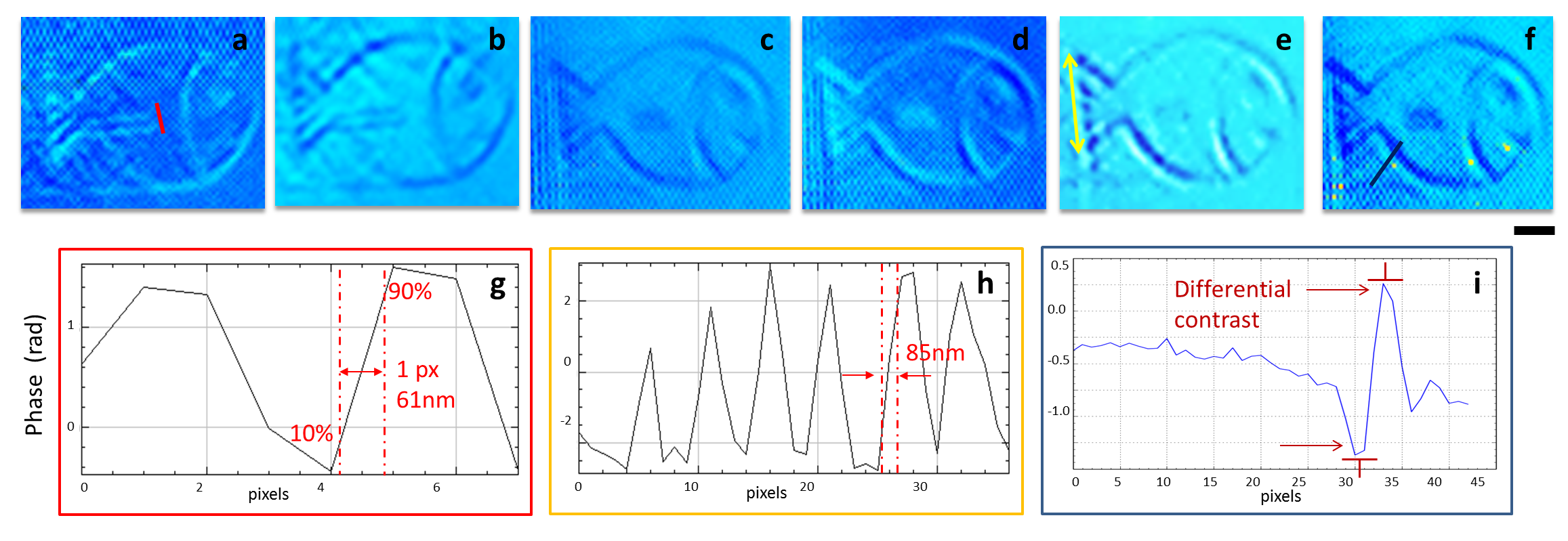}
	\caption{{\small (a-f) Phase reconstruction filtered by background subtraction and scaled intensity at 500 sec (a,b,e,f) and 200 sec(c,d) exposure; g. knife-edge resolution test of red line profile in a; (h,i) profile plot of yellow and black line in e,f respectively.}}
	\label{recs}
\end{figure*}
\\
\indent To demonstrate the superior performance of the iridium/HSQ nano-wireframe reference by ALD fabrication compared to the bare silicon pillar by basic lithographic etching, we study the intensity distribution of holographic measurements obtained with the two references.
Figure \ref{refmod}(a) shows a plot of the intensity distribution measured with the ALD reference (pink and black curve), overplotted with data acquired using the bare silicon pillar at varying exposure time (red,blue, and green curves). There is a clear increase in fringe visibility as well as an evident increase in resolution for the ALD data. With the well-resolved outermost fringes of the ALD data reaching the edge of the detector, the experimental resolution reaches its maximum limit of 1 pixel size, i.e 61 nm at 9.1 keV.  We can safely conclude that the resolution can reach even lower values if the detector area is extended beyond 515 pixels. In comparison, the 200sec exposure fringes of the bare silicon reference (red curve) overall have less visibility and begin to be distorted beyond the 220 pixel mark, resulting in a resolution  limited by the reference quality and not the detector area.\\
\indent By inspecting the reference autocorrelation we are able to characterise the structural attributes of the reference. The autocorrelation representing four adjacent complex-conjugate copies of the reference structure is shown in Figure \ref{refmod}(b). From these images we identify the iridium nano-layer as a sharp, thin, and intense straight line enveloping the underlying HSQ structure which has lower intensity and visibly rough edge corrugations (Figure \ref{refmod}(c)). Since the HSQ pillar on its own is fabricated in a similar fashion to the previously discussed bare silicon pillar, Figure \ref{refmod}(c) directly visualises the difference in the resulting structure of each fabrication method. The profile plot (Figure \ref{refmod}(d)) across a single quadrant shows a sharp peak compared to the almost fully transmissive inner area. This demonstrates that despite the 10nm width of the iridium nano-layer, the $3.9\mu m$ height is effectively absorbing, hence the high contrast reconstructions. In addition, the profile plot across the reference edge (Figure \ref{refmod}(e)) shows the sharp iridium peak adjacent to the broad peak corresponding to the rough HSQ edge. The narrowness and sharp tip of the iridium peak illustrates the enhanced imaging resolution of the ALD reference, backed by the distribution of fringes reaching the detector limit in Figure \ref{refmod}(a) compared to that of the basic reference pillar. Characterising the reference structure helps us to understand some anomalies that arise in the sample reconstructions that cannot be attributed to computational reasons. For example, the appearance of faint repeating copies of the sample reconstruction (in the left side of Figure \ref{schem}(f)) is attributed to the pillar edge corrugations, where each sharp corrugation acts as an individual holographic point source reference. The fact that the repeating copies have a weak signal and low resolution validates them being a cause of single broad point references of the low-phase/low-absorption HSQ structure. Moreover, the fact that such copies do not appear at symmetrical steps with respect to one another or at all sides of the virtual reference justifies our conclusion.

For the given detector size and photon energy, we demonstrate the full capability of the imaging system in the high-resolution reconstructions shown in Figure \ref{recs}. This is accomplished by refining the illumination beam profile, increasing the incident flux, and applying data correction by subtracting the individual autocorrelations of the beam and reference from the holographic autocorrelation array. 
Reconstructions from data acquired at a 500sec exposure time are shown in Figure \ref{recs}(a,b) with an accumulated $3.89 \times 10^{10} $ photons, where the finest features of the sample are now resolvable i.e the two fins of the fish with the sub-50 nm gap in between them. Furthermore, the profile plot in Figure \ref{recs}(h) across the tail structure of Figure \ref{recs}(e) shows the high contrast of the resolved detail. By calculating the 90$\%$-10$\%$ knife-edge plots of the sample features (Figure\ref{recs}(g)), we find that the holographic imaging system using the ALD iridium wireframe reference has reached the pixel-size resolution limit of 61 nm, i.e the maximum attainable image resolution for the given photon energy and pixel size.
To further the resolution limit, one could increase the photon energy or utilise a larger detector array to effectively decrease the object-space pixel size and to capture the maximum scattering angle. The virtue of reference structures prepared by atomic layer deposition is that the resolution could be enhanced by preparing an ultra-sharp layer of a few nanometer or even sub-nanometer thickness, without affecting the SNR or contrast.
\begin{figure*}[!t]
	\vspace{-0.3cm}
	\includegraphics[scale=0.42]{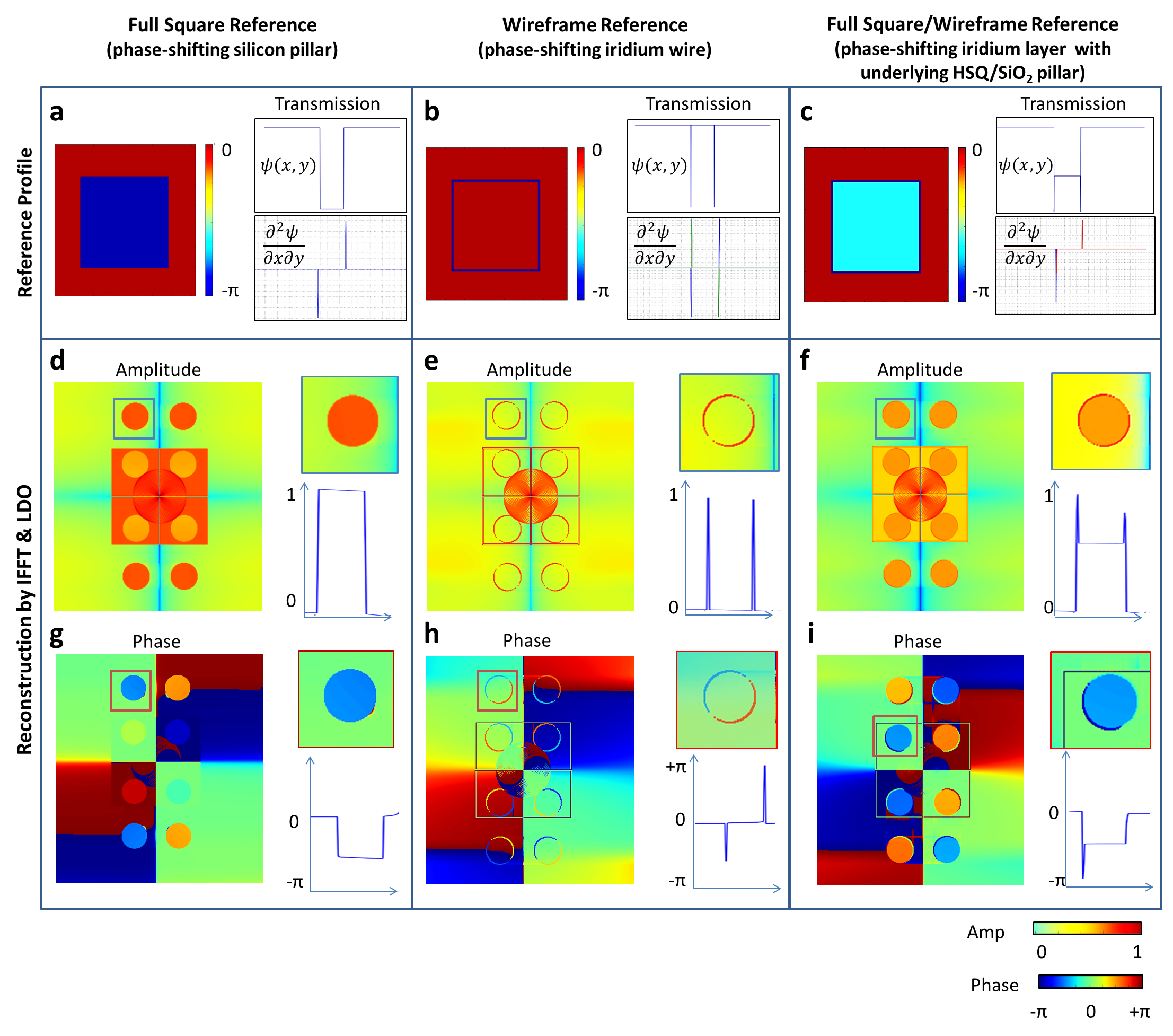}
	\caption{{\small Simulation comparison of three reference types and their amplitude and phase reconstructions: of (a,d,g) full square, (b,e,h) wireframe, and (c,f,i) full-square/wireframe reference, with respective plot profiles of magnified cropped areas.}}
	\label{sim}
\end{figure*} 

There is a unique attribute common to all the reconstructions shown in Figure \ref{schem} and  \ref{recs} that does not appear for reconstructions obtained with the basic silicon pillar or for results from previous FTH and HERALDO methods. That is the differential contrast and edge contrast enhancement. As evident in both the phase and amplitude reconstructions in Figure \ref{schem}(e) and (f) as well as the corrected phase reconstructions in Figure \ref{recs}(a-e), the intensity specifically at the outline of the fish sample alternates from high to low across a single feature, which gives the effect of enhanced edge contrast. 
For example, in the magnified reconstructions in Figure \ref{schem}(e) and (d), the signal of the upper left outline goes from bright (outer outline) to dark (inner outline); and the contrast propagates diagonally across to the lower right side of the fish. This is illustrated in the profile plot in Figure\ref{recs}(i) where the outline signal constitutes a negative dip followed by a positive peak. The differential phase contrast is a direct consequence of the wireframe reference geometry and the resulting arrangement of double delta peaks with opposite polarity at a single corner. The differential phase contrast is a function of the phase difference between the iridium layer and the underlying structure. 
The diagonal direction flips alternatingly between each reconstruction at the four virtual corners, due to the alternating arrangement of the double delta peaks.

\vspace{-3pt}
The fact that the reference structure is a wireframe implies that the holographic mechanism is not equal to that of full-area square reference. In order to understand this effect, we must return to the mathematical representation of the complex transmissivity of the reference function and the consequence of applying the linear differential operator. This is better illustrated in the simulations in Figure \ref{sim} that compare three types of reference geometry: a full-area square (Figure \ref{sim}(a)), a square wireframe (Figure \ref{sim}(b)), and a wireframe with an underlying square structure (Figure \ref{sim}(c)), neglecting substrate contribution and assuming a homogeneous parallel incident beam. The sample represents a 340nm-thick gold disc placed on a transmissive substrate, analogous to our experiment. The parameters are chosen to mimic the experimental data of a 9$\mu m$ silicon pillar, a 3.9$\mu m$ iridium frame, and a 3.9$\mu m$ HSQ/iridium structure, where the silicon and iridium are both $\pi$ phase-shifting.  
The delta peaks resulting from the double directional derivative $\partial^2 \psi / \partial x\partial y$ of the reference transmission function are shown in the profile plots in Figure \ref{sim}(a,b,c) along a single side. 
The reconstructions produced by the full-area square reference show a flat intensity of amplitude and phase with no outline contrast (Figure \ref{sim}(d,g)). For the pure wireframe case (Figure \ref{sim}(b,e,h)), the reconstructions are merely outline images with no bulk information (yet for thicker samples, the phase includes the bulk information). In Figure \ref{sim}(h), the phase of the outline switches from negative to positive in the diagonal direction. For a single wireframe, applying a double directional derivative to the transmission function shown in Figure \ref{sim}(b)(top) results in the derivative of a delta function, i.e two concurrent peaks of opposite polarity. The two opposite peaks induce the contrast gradient across the outlines. In the case of the wireframe reference with an underlying pillar of HSQ (Figure \ref{sim}(c)), we find a reconstructed full-area image of the gold disc as well as enhanced outline contrast (Figure \ref{sim}(f)) and a phase contrast gradient (Figure \ref{sim}(i)). In our experimental setup, the underlying HSQ pillar was originally designed for facilitated fabrication of an ALD wireframe layer. Yet, the results prove that reconstructions produced with such a reference preserve the bulk information, owing to the underlying pillar, and portray additional edge contrast enhancement both in amplitude and phase, owing to the high absorption and $\pi$ phase shift of the iridium nano-layer. The differential contrast can be particularly useful for imaging applications with weakly scattering or low density samples where the effect of edge enhancement increases detectability and heightens the contrast with respect to the background signal. Another application is for specific samples where the focus of investigation is in the outline and boundary details such as cell membranes and semiconductor interfaces.

\indent In conclusion, in this paper we have demonstrated that the holographic method employing the unique phase-shifting 3D reference design results in high-resolution image reconstructions with differential phase contrast as well as edge enhancement. The method is noise-robust due to the geometrical shape of the reference, it circumvents the resolution-contrast trade-off inherent to single point references and surpasses the limitations of holographic methods associated with absorbing reference masks. By successfully implementing high-resolution holographic imaging with hard X-rays, the method is no longer confined to applications with soft x-rays but it bridges over to high-energy coherent radiation. The method is easily applicable to thicker, more dense, and larger samples such as bone tissue, and the field of view can be increased accordingly by larger lateral reference dimensions. The unconfined experimental setup enhances the measurement quality, permits live feedback, provides means for effective background filtering. The enabled independent sample translation and rotation allow data averaging as well as multiple-sample data acquisition in a single experiment, and introduces new possibilities for tomographic imaging \cite{tomo}.  Single-shot data can be reconstructed to produce absorption and phase images by simple computation without the use of iterative algorithms. The refinement of the reference structure by atomic layer deposition and nano-fabrication advancements holds potential to boost resolution to the single nanometer range. The full capacity of the holographic method can be further exploited by employing large-area detectors, smaller pixel size, and higher flux. With all these advantages and more, such a non-iterative high-resolution imaging method can greatly benefit from the highly brilliant and coherent sources available at third-generation synchrotrons and proves highly applicable to single-shot femtosecond imaging of biological samples \cite{biopulsesneutze,biowindow} and ultrafast processes \cite{femto1fthferro,femtoseq} using x-ray free-electron-lasers \cite{chapxfelsoft,spencediffdest}.
\\
% acknowledgement.tex                            1 December 2015
%
% Acknowledgement paragraph with agency abbreviations of Oct. 15, 2014 for non-APS journals

We would like to thank the University of Z{\"u}rich for financial funding. We thank the beamline management of I13 and the user office at Diamond Light Source for providing the beamline where all the experimental work was conducted and covering subsistence costs during the beam-time. We thank the Laboratory for Micro- and Nanotechnology lead at the Paul Scherrer Insitute.   % input acknowledgement


\begin{thebibliography}{99}
{\small \bibitem{chapmanprotein}
H. N. Chapman {\sl et al.}, Nature {\bf 470}, 73-77 (2011).
\bibitem{spencediffdest}
J. Spence, Nat. Photonics {\bf 2} (7), 390-391, (2008).
\bibitem{mimivirus}
M. M. Seibert {\sl et al.}, Nature 470, 78-81 (2011).
\bibitem{chapxfelsoft}
H. N. Chapman {\sl et al.}, Nat. Phys. {\bf 2}, 839-843 (2006).

\bibitem{biopulsesneutze}
R. Neutze, Nature {\bf 406}, 752-757 (2000).

\bibitem{gabor}
D. Gabor {\sl et al.}, Nature {\bf 161}, 777-778 (1948).
\bibitem{mcnulty}
I. McNulty {\sl et al.}, Science {\bf 256}, 1009-1012 (1992).
\bibitem{eisebit1}
S. Eisebitt {\sl et al.}, Nature {\bf 432}, 885-888 (2004).
\bibitem{guizarmain}
M. Guizar-Sicairos and J. R. Fienup, Opt. Express {\bf 15}, 17592-17612 (2007).
\bibitem{heraldozhucheese}
D. Zhu {\sl et al.}, Phys. Rev. Lett. {\bf 105}, 043902 (2010).
\bibitem{multsoftfth}
W. F. Schlotter {\sl et al.}, Appl. Phys. Lett {\bf 89}, 163112 (2006).

\bibitem{manuthesis}
M. Guizar-Sicairos, Ph.D Thesis, University of Rochester, New York, (2008).
\bibitem{URA}
S. Marchesini {\sl et al.}, Nat. Photonics {\bf 2}, 560-563 (2008).
\bibitem{ilruft}
A. V. Martin {\sl et al.}, Nat. Comms. {\bf 5}:4661, (2014).
\bibitem{podorov}
S. G. Podorov {\sl et al.}, Opt. Express {\bf 15} (16), 9954-9962 (2007).
\bibitem{podorovmask}
S. G. Podorov {\sl et al.}, Ultramicroscopy {\bf 111} (7) 782-787 (2011).

\bibitem{podorov3d}
S. G. Podorov {\sl et al.}, Applied Optics {\bf 55} (3), A150-A153 (2016). 

\bibitem{tomo}
E. Guehrs {\sl et al.}, %Mask-based dual-axes tomoholography using soft x-rays
, New J. Phys {\bf 17}, 103042 (2015).
\bibitem{femtoseq}
C. M. Gunther {\sl et al.}, Nat. Photonics {\bf 5}, 99-102 (2011).
\bibitem{heraldofemto}
D. Gauthier {\sl et al.}, Phys. Rev. Lett. {\bf  105}, 093901 (2010).
\bibitem{heraldointegralcrosswire}
M. Guizar-Sicairos {\sl et al.}, Opt. Lett. {\bf 35}(7) (2010).
\bibitem{heraldowindocapo}
F. Capotondi {\sl et al.}, Opt. Express {\bf  20}, 25152-25160 (2012).
\bibitem{hard1}
W. Leitenberger {\sl et al.}, Proc. VI Int. Conf. X-ray Microscopy, 507:497 (2000).

\bibitem{hard2}
L. M. Stadler {\sl et al.}, Phys. Rev. Lett. {\bf 100}, 245503 (2008).

\bibitem{enders}
B. Enders {\sl et al.}, New J. Phys. {\bf 11}, 043021 (2009).
\bibitem{enders2}
B. Enders {\sl et al.}, Diplomarbeit, Georg-August-Universit{\"a}t zu G{\"o}ttingen, (2009).

%27

\bibitem{mirna}
M. Saliba {\sl et al.}, manuscript in preparation (2016).
\bibitem{monolithic}
J. Geilhufe {\sl et al.}, Nat. Commun. {\bf 5}, 3008 (2014).


\bibitem{biowindow}
T. ~Gorniak {\sl et al.}, Opt. Express {\bf 19}, 11059-11070 (2011).

\bibitem{femto1fthferro}
T. Wang {\sl et al.}, Phys Rev Lett {\bf 108}, 267403 (2012).
}

%-------------------------------------------
%\bibitem{chapmanprotein}
%H. N. Chapman {\sl et al.}, Femtosecond x-ray protein nanocrystallography, Nature 470, 73-77 (2011).
%
%\bibitem{spencediffdest}
%J. Spence, X-ray imaging: Ultrafast diffract-and-destroy movies, Nat. Photonics 2(7), 390-391, (2008).
%\bibitem{mimivirus}
%M. M. Seibert {\sl et al.}, Single mimivirus particles intercepted and imaged with an x-ray laser, Nature 470, 78-81 (2011).
%\bibitem{biopulsesneutze}
%R. Neutze, Potential for biomolecular imaging with femtosecond X-ray pulses, Nature 406, 752-757 (2000).

%\bibitem{chapxfelsoft}
%H. N. Chapman {\sl et al.}, Femtosecond diffraction imaging with a soft X-ray free-electron laser, Nat. Phys. 2, 839-843 (2006).
%\bibitem{gabor}
%D. Gabor {\sl et al.}, A new microscopic principle, Nature 161, 777-778 (1948).
%\bibitem{mcnulty}
%I. McNulty {\sl et al.}, High-resolution imaging by Fourier transform X-ray holography, Science 256, 1009-1012 (1992).
%\bibitem{eisebit1}
%S. Eisebitt {\sl et al.}, Lensless imaging of magnetic nanostructures by X-ray spectro-holography, Nature 432, 885-888 (2004).
%\bibitem{guizarmain}
%M. Guizar-Sicairos and J. R. Fienup, Holography with extended reference by autocorrelation linear differential operation, Opt. Express 15, 17592-17612 (2007).
%\bibitem{heraldozhucheese}
%D. Zhu {\sl et al.}, High-resolution x-ray lensless imaging by differential holographic encoding, Phys. Rev. Lett. 105, 043902 (2010).
%\bibitem{multsoftfth}
%W. F. Schlotter {\sl et al.}, Multiple reference fourier transform holography with soft x-rays, Appl. Phys. Lett 89, 163112 (2006).
%
%\bibitem{manuthesis}
%M. Guizar-Sicairos, Methods for coherent lensless imaging and x-ray wavefront measurement, Ph.D Thesis, University of Rochester, New York, (2008).

%\bibitem{URA}
%S. Marchesini {\sl et al.}, Massively parallel x-ray holography, Nat. Photonics 2, 560-563 (2008).
%\bibitem{ilruft}
%A. V. Martin {\sl et al.}, X-ray holography with a customizable reference, Nat. Comms. 5:4661, (2014).
%\bibitem{femtoseq}
%C. M. Gunther {\sl et al.}, Sequential femtosecond X-ray imaging, Nat. Photonics 5, 99-102 (2011).
%\bibitem{heraldofemto}
%D. Gauthier {\sl et al.}, Single-shot femtosecond x-ray holography using extended references, Phys. Rev. Lett. 105, 093901 (2010).
%\bibitem{heraldointegralcrosswire}
%M. Guizar-Sicairos {\sl et al.}, Holographic x-ray image reconstruction through application of differential and integral operators, Optics Letters 35(7) (2010).
%\bibitem{heraldowindocapo}
%F. Capotondi, A scheme for lensless X-ray microscopy combining coherent diffraction imaging and differential corner holography, Opt. Express 20, 25152-25160 (2012).
%\bibitem{hard1}
%W. Leitenberger, Fourier transform holography with coherent hard x-rays, Proc. VI Int. Conf. X-ray Microscopy, 507:497 (2000).
%
%\bibitem{hard2}
%L. M. Stadler, Hard X-ray holographic diffraction imaging, Phys. Rev. Lett. 100, 245503 (2008).
%
%\bibitem{podorov}
%S. G. Podorov {\sl et al.}, A non-iterative reconstruction method for direct and unambiguous coherent diffractive imaging, Opt. Express 15, 9954-9962 (2007).
%\bibitem{enders}
%B. Enders {\sl et al.}, Non-iterative coherent diffractive imaging using a phase-shifting reference frame, New J. Phys. 11, 043021 (2009).
%\bibitem{enders2}
%B. Enders {\sl et al.}, Single-step inversion for coherent diffraction imaging - a generalisation to fourier transform holography, Diplomarbeit, Georg-August-Universit{\"a}t zu G{\"o}ttingen, (2009).

%
%\bibitem{monolithic}
%J. Geilhufe {\sl et al.}, Monolithic focused reference beam x-ray holography, Nat. Commun. 5, 3008 (2014).
%\bibitem{ald}
%REF ALD fabrication method description.
%%27
%\bibitem{litho}
%REF E-beam lithography
%\bibitem{mirna}
%M. Saliba {\sl et al.}, Hard X-ray Holographic Diffraction Imaging Using 3D Phase-Shifting References in a Deconfined Setup, manuscript in preparation (2016).
%\bibitem{merlin}
%REF Merlin Detector
%
%\bibitem{highald}
%High resolution ALD fabrication
%
%\bibitem{biowindow}
%T. ~Gorniak {\sl et al.}, X-ray holographic microscopy with zone plates applied to biological samples in the water window using 3rd harmonic radiation from the free-electron laser FLASH, Opt. Express {\bf 19}, 11059-11070 (2011).
%
%\bibitem{femto1fthferro}
%T. Wang {\sl et al.}, Femtosecond single-shot imaging of nano-scale ferromagnetic order in Co=Pd multilayers using resonant x-ray holography, Phys Rev Lett 108, 267403 (2012).


\end{thebibliography}
\end{document}